\documentclass[prb,reprint,twocolumn,superscriptaddress,noshowpacs,notitlepage,longbibliography,10pt,citeautoscript]{revtex4-1}%
\pdfoutput=1
\usepackage{graphicx}% Include figure files
\usepackage{amsmath,amssymb}
\usepackage{enumitem} % Customized lists
\usepackage{mathtools}
\usepackage{afterpage}
\usepackage{epigraph}
\usepackage{hyperref}
\hypersetup{
	colorlinks = true,
	allcolors = {blue}
}
\usepackage{cleveref}
\usepackage[title]{appendix}
\usepackage{bm}% bold math
\DeclarePairedDelimiter\bra{\langle}{\rvert}
\DeclarePairedDelimiter\ket{\lvert}{\rangle}
\DeclarePairedDelimiterX\braket[2]{\langle}{\rangle}{#1 \delimsize\vert #2}

\begin{document}

\title{Composite Topological Weyl Nodal lines}

\author{Oliver Dowinton}
\email{oliver.dowinton@postgrad.manchester.ac.uk}
\affiliation{Department of Physics and Astronomy, The University of Manchester, Oxford Road, Manchester M13 9PL, United Kingdom}
\author{Rodion Vladimirovich Belosludov}
\affiliation{Institute for Materials Research, Tohoku University, Sendai 980-08577 Japan}
\author{Mohammad Saeed Bahramy}
\email{m.saeed.bahramy@manchester.ac.uk}
\affiliation{Department of Physics and Astronomy, The University of Manchester, Oxford Road, Manchester M13 9PL, United Kingdom}

\date{\today}
\vspace{3mm}

\begin{abstract}
Nodal lines are one-dimensional topological features of semi-metal band structures along which two bands are degenerate as a result of non-{\it accidental} symmetry-protected crossings, and behave topologically as $k$-space vortices in the Berry connection. Here, we present a new class of tilted nodal lines, protected by mirror symmetry, formed from the intersection of three band crossings at a set of critical points. One crossing is gapped out, fusing the remaining two crossings at the shifted critical points to form composite nodal lines.  We demonstrate these composite nodal lines are capable of supporting fundamentally different Berry curvature textures than the typical two-band case, despite having a simple ring topology. In addition, we present a realistic model based on cubic, forced-ferromagnetic, EuTiO$_3$, where the spin and orbital degrees of freedom are plentiful enough to allow the material realization of such composite nodal lines. In this system, the composite nature of the nodal line results in a spin Hall conductivity with a non-monotonic dependence on carrier concentration.
\end{abstract}
\maketitle
\section{Introduction}

In a solid-state system, as two bands approach each other energetically they will generically hybridize, resulting in an energy gap between them.
However, if the system possesses certain topological or symmetry constraints, the two bands can cross without any hybridization, leading to a perfectly protected degenerate band crossing.
Such crossings are the defining feature of topological semi-metals (TSMs) and take the form of discrete points (Weyl or Dirac nodes)~\cite{Murakami_2007,Young_2012,Burkov_2011_point,Wan_2011,Xu_2015,Weng_2015,Huang} or one-dimensional (1D) manifolds (nodal lines)~\cite{Burkov_2011_line,Rauch_2017,Fang_2016,Heikkila_2015,Xu_2011,Weng_2015,Huang}. Naively, the nodal lines must form closed loops in the Brillouin zone (BZ). However, by analogy to the behaviour of topologically charged vortices in high energy physics and superfluids, intersecting nodal lines may `annihilate' each other, closing at discrete points known as nexuses~\cite{Heikkila_2015,Hyart_2016,Chang_2017,Lenggenhager_2022,Chen_2018}. 

Whereas previous realizations of nodal line nexuses/triple degeneracy points~\cite{Hyart_2016,Chang_2017,Lenggenhager_2022,Chen_2018} typically use the intersection of multiple symmetry planes and hence nodal line topologies more complex than a ring, we examine a simplified toy model of a Weyl nodal line where three crossings meet at critical points within a single mirror plane. 
Introducing additional terms that respect the mirror symmetry gaps one crossing out, leaving remaining nodal lines that, while having a simple ring topology, can be viewed as the composite of two nodal lines that meet and annihilate at the shifted critical points, i.e. nexuses.
To explore this further, some clarification is made on the Berry curvature (BC) textures from typical nodal lines, allowing us to demonstrate that, when having a small gap, this composite nodal line is capable of BC textures fundamentally different from the standard two-state case due to the presence of the third band, with the two components of each composite nodal line having opposite chirality, and the nexuses acting like sources or sinks of BC.

For a realistic material proposal, we examine density functional theory (DFT) calculations and a simple tight-binding model for the forced-ferromagnetic(FFM) cubic phase of EuTiO$_3$, a system where anomalous Hall conductivity has been connected to Weyl semi-metal topology~\cite{Takahashi_2018,Ahadi2018,Dowinton_2022}. The tight-binding model is based on a conduction manifold formed of $d$-$t_{2g}:\{d_{xy},d_{xz},d_{yz}\}$ orbitals subject to a mirror plane symmetry. We demonstrate that this orbital manifold is ideal for hosting composite nodal lines, hierarchically formed by an initial crossing between bands with like spins, which then evolve into bands with opposite spins. This configuration results in spin Hall conductivity (SHC) exhibiting a characteristic turning point behavior in its energy (carrier density) dependence at the location of the nexus.

\section{Mirror Symmetry Protected Nodal Line Toy Models}
%FIGURE1--------------
\begin{figure}
    \includegraphics[width=\linewidth]{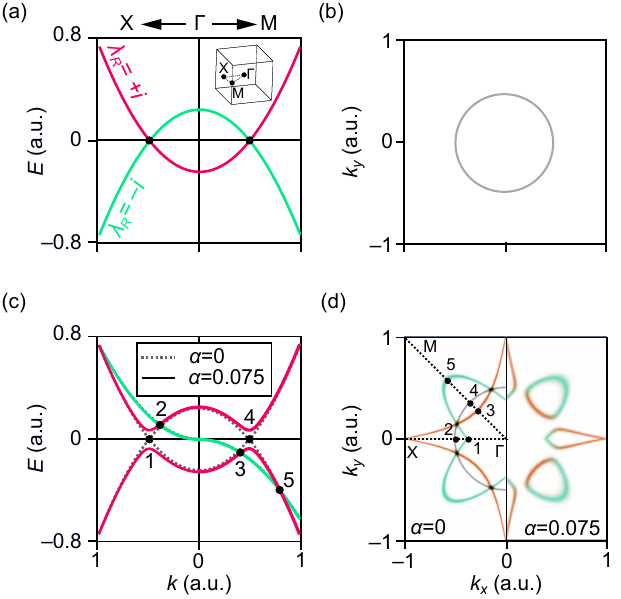}
    \caption{Schematic overview of mirror-plane protected nodal lines.
    (a-b) Electronic structure for circular toy model with radius $\rho=0.5$.
    (a) Band structure along $M \rightarrow \Gamma \rightarrow X$ high symmetry lines.
    Color coding shows positive and negative mirror eigenstates in pink and cyan respectively.
    (b) Nodal line shape in  $k_z =0$ plane.
    (c-d) Electronic structure for 3 band toy model with $\mu =0.75, \tau=2.75, \rho=0.5$.
    (c) Band structure along $X \rightarrow \Gamma \rightarrow M$ high symmetry lines, with the dashed lines showing the $\alpha =0$ case, and the solid lines showing the $\alpha =0.075$ case.
    The dots mark the intersection of the crossings with the high symmetry lines. (d) Nodal lines for 3 band model in $k_z =0$ plane, with `star' nodal line in red, `flower' in cyan, and circular in grey. The left-hand side shows shapes of the crossings for $\alpha =0$, with corresponding dots from (c). The right-hand side shows composite nodal lines for $\alpha=0.075$.}
    \label{fig:1}
\end{figure}
In a three-dimensional (3D) BZ, 1D defects, {\it i.e.} nodal lines, are not true topological defects but require additional symmetry protections to be stable~\cite{Burkov_2011_line,Horava}. While a variety of symmetries can stabilise nodal lines~\cite{Fang_2016}, this paper only deals with Weyl nodal lines protected by mirror symmetries~\cite{Fang_2016,Xu_2011}.
In addition, while previous studies have focused on nodal line nexuses from the intersection of multiple mirror planes~\cite{Hyart_2016,Chang_2017,Lenggenhager_2022,Chen_2018}, we will discuss a three-band system with intersecting nodal lines in simply one mirror symmetry plane, which without loss of generality is taken to be the $xy$ plane, {\it i.e.} the lattice is invariant under a reflection that takes $(x,y,z) \rightarrow (x,y,-z)$.

The presence of mirror symmetry allows stable nodal lines because in $k$-space, the mirror operator sends the $k_z=0,\pi$ planes to themselves, such that any energy-eigenstates must also be eigenstates of the mirror operator. Labelling the bands by their mirror eigenvalue $\lambda_R = \pm i$, bands, $n,m$, may form protected crossings if $\lambda_R^{(n)}\neq \lambda_R^{(m)}$, as they are incapable of mixing due to the mirror symmetry.

The model considered begins as an extension of the typical circular nodal line toy model~\cite{Bian_2016,Chari_2016,Ekstrom_2021}
\begin{equation}
    \mathcal{H}(\bm{k})= (k_x^2+k_y^2-\rho^2) \sigma_z + \beta k_z \sigma_y,
\end{equation}
where $\rho,\beta > 0$, and $\bm{\sigma}$ are the Pauli matrices.
This is constructed to have a mirror symmetry with respect to $R = i \sigma_z$.
$\mathcal{H}(\bm{k})$ has eigenvalues
\begin{equation}
    \varepsilon_\pm = \pm\sqrt{(k_x^2+k_y^2-\rho^2)^2+k_z^2\beta^2},
\end{equation}
with the circularly symmetric band structure in the $k_z=0$ plane shown in Fig.~\ref{fig:1}(a), color coded with mirror eigenvalues.
These bands are degenerate when $\varepsilon_\pm=0$, {\it i.e.} when $k_z =0$ and $k_x^2+k_z^2 = \rho^2$. This results in a circular nodal line with radius $\rho$ in the $k_z=0$ mirror invariant plane, Fig.~\ref{fig:1}(b), which is protected by mirror symmetry as any perturbation that respects mirror symmetry cannot hybridise the $\sigma_z$ eigenstates. 

Extending this to three bands, there are two scenarios: either all the bands have the same mirror eigenvalues in the $k_z = 0$ plane, such that the symmetry does not confer any protection to crossings, or two bands have equal mirror eigenvalues, and the third has the opposite, allowing the formation of protected nodal line structures in the $k_z = 0$ plane.

Considering the latter case, we examine a toy model system described by the Hamiltonian:
\begin{widetext}
\begin{equation*}
    \mathcal{H}\left( \bm{k} \right) = \begin{pmatrix}
        k_x^2 +k_y^2 - \rho^2 & - i \beta k_z & i \alpha - i \beta k_z \\
        i \beta k_z & \mu (k_x^2 +k_y^2) - \tau \sqrt{k_x^2 k_y^2} & -i\beta k_z\\
        - i\alpha +i \beta k_z  & i\beta k_z & -k_x^2 -k_y^2 + \rho^2
    \end{pmatrix},
\end{equation*}
\end{widetext}
where the mirror operator in this basis is given by:
\begin{equation}
    \mathcal{R} = \begin{pmatrix}
        1 & 0 & 0\\
        0 & -1 & 0\\
        0 & 0 & 1 
    \end{pmatrix}.
\end{equation}
The band-structure along $X\rightarrow\Gamma\rightarrow M$ high symmetry lines is shown in Fig.~\ref{fig:1}(c).
Initially, taking $\alpha =0$ implies further symmetry restrictions than the mirror symmetry alone by removing the mixing between the bands that have equal mirror eigenvalues. In this case, there are three perfect crossings in the $k_z=0$ plane, defined implicitly by the equations
\begin{equation}
\begin{split}
    k_x^2+k_y^2-\rho^2&=0\\
    (\mu \pm 1)(k_x^2+k_y^2) \mp \rho^2 -\tau\sqrt{k_x^2 k_y^2}&=0,
\end{split}
\end{equation}
where the first is the circular nodal line, and the second describes the `star' and `flower' shaped nodal lines, shown in the left-hand side of Fig.~\ref{fig:1}(d).
These all intersect at a set of eight triply degenerate critical points given by
\begin{equation}
\begin{split}
    k_{0,x}^2&=\frac{1}{2}\left(\rho^2\pm\sqrt{\rho^4-4\frac{\mu^2 \rho^4}{\tau^2}}\right)\\
    k_{0,y}^2&=\frac{1}{2}\left(\rho^2\mp\sqrt{\rho^4-4\frac{\mu^2 \rho^4}{\tau^2}}\right).\\
\end{split}
\end{equation}

Increasing $\alpha$ introduces a gap to the circular crossing with radius $\rho$, Fig.~\ref{fig:1}(c), reconstructing the remaining protected crossings as shown in the right-hand side Fig.~\ref{fig:1}(d), where they are composed from parts of the original star/flower shaped nodal lines `fused' at the intersections with the circular avoided crossing, {\it i.e.} at a set of 16 nexus points given by
\begin{equation}
\begin{split}
    k_{0,x}^2&=\frac{1}{2}\left(\rho^2\pm\sqrt{\rho^4-4\left(\frac{\mu \rho^2+\gamma \alpha}{\tau} \right)^2}\right)\\
    k_{0,y}^2&=\frac{1}{2}\left(\rho^2\mp\sqrt{\rho^4-4\left(\frac{\mu \rho^2+\gamma \alpha}{\tau} \right)^2}\right),
\end{split}
\end{equation}
where $\gamma =\pm 1$ for either of the two shapes of the remaining nodal lines.

\section{Classifying Nodal Line Chirality with Berry Curvature Texture}
%FIGURE2--------------
\begin{figure*}[ht]
    \includegraphics[width=\linewidth]{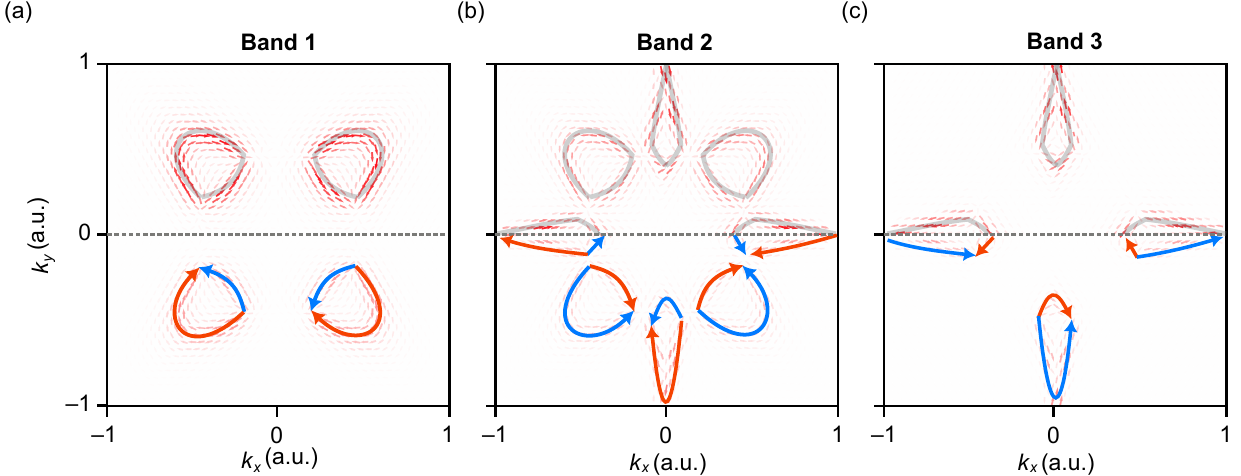}
    \caption{Band-resolved contributions to the BC for a mirror symmetry breaking gap $\gamma = 0.001$ eV. The bottom half shows the direction of the BC flow schematically.}
    \label{fig:2}
\end{figure*}
This section deals with the topology and Berryology of the composite nodal lines, with a view to connecting it to previous arguments about nexuses behaving like \mbox{(anti-)monopoles}, and nodal lines like vortex flux lines in nexus nodal line semi-metals~\cite{Heikkila_2015}.

 Nodal rings are symmetry-protected topological features because integrating the Berry connection around a contour, $\mathcal{C}$, that encloses the nodal line results in a $\varphi=\pi$ Berry phase, insensitive to the specific geometry of $\mathcal{C}$~\cite{Berry1984,Burkov_2011_line,Fang_2016,Chari_2016,Bian_2016,Li_2018},
\begin{equation}
\begin{split}
    \varphi &= \int_{\bm{\mathcal{C}}} \bm{\mathcal{A}} \cdot d\bm{l}=\pi\\
     \bm{\mathcal{A}}(\bm{k}) &= i\sum_n f_n(\bm{k}) \braket{n}{\bm{\nabla}_{\bm{k}} n},
\end{split}
 \end{equation}
 where $f_n(T,\bm{k})$ is the Fermi-Dirac distribution.
 Hence, one can view the nodal line as a $k$-space vortex in the Berry connection.

 Furthermore, by naively invoking Stokes' theorem
 \begin{equation}
 \varphi= \int_{\bm{\mathcal{C}}} \bm{\mathcal{A}} \cdot d\bm{l} = \int_{\bm{\mathcal{S}}} \bm{\Omega} \cdot d\bm{\mathcal{S}},
 \end{equation}
 where $\bm{\Omega}$ is the Berry curvature (BC),
 \begin{equation}
    \begin{split}
        \Omega_{ij}(\bm{k})&=\sum_n f_n (T,\bm{k}) \omega_{n,ij}(\bm{k})\\
        \omega_{n,ij}&=-\hbar^2\mathfrak{Im}\sum_{m\neq n}\frac{\bra{n}\hat{v}_i\ket{m}\bra{m}\hat{v}_j\ket{n}}{(\varepsilon_n-\varepsilon_m)^2}\\
        \hat{v}_i&=\frac{1}{\hbar} \frac{\partial \widehat{\mathcal{H}}}{\partial k_i},
    \end{split}
\end{equation}
 one could be tempted to view the nodal lines as flux tubes of quantized $\bm{\omega_\pm}\cdot\hat{\bm{n}} =\pm \pi$ BC contributions from each band that circulate around the nodal line, where $\hat{\bm{n}}$ is the normal of $\mathcal{S}$, and are zero everywhere else. Thus extending the analogy of vortex flux line nexuses, we could wonder if our composite nodal lines behave like fused flux lines with different chirality of BC, with nexuses acting as sources and sinks of BC.

However, this picture misses some subtleties because while a Berry phase is demonstrable, inferring the BC from Stokes' theorem is mathematically dubious. 
In the case of a zero-dimensional Weyl node, the BC contribution from each band can be computed by considering the flux through a $k$-space surface, e.g. sphere, that does not contact the node, unambiguously assigning a charge of $\pm \pi$ as in Berry's original paper~\cite{Berry1984}. 
Whereas for the extended 1D nodal line any 2D surface used to compute the BC will contact the nodal line where the BC is singular.
One could attempt to circumvent this by adding an arbitrary parameter $R$ to the system, allowing the existence of a surface in $(\bm{k},R)$-space where the nodal line behaves like a 0D node such that a BC in the original $k$-space can be assigned. However, the choice of $R$ is a gauge freedom that the BC is sensitive to.
This gauge freedom is anticipated in the original motivating argument, noting that the Berry phase is defined modulo $2\pi$; hence, $\pm \pi$ phases are physically equivalent such that the vorticity of the Berry connections is gauge dependent.

To proceed, it is necessary to consider a small perturbative gap that allows us to define a BC texture for the nodal line. While this has been done for simple nodal line models, for example, in Li et al.~\cite{Li_2018} Yang et al.~\cite{Yang_2022}, it is critical to rigorously show that any novel result for the three-band case is fundamentally distinct from a generic two-band case.
Hence, generically, a two-band nodal line has a Hamiltonian of the form:
\begin{equation}
    \mathcal{H}(\bm{k}) = \varepsilon_0 (\bm{k}) \sigma_0 + h_x(\bm{k}) \sigma_x + h_z(\bm{k}) \sigma_z,
\end{equation}
with energy eigenvalues
\begin{equation}
    \varepsilon_\pm(\bm{k}) = \varepsilon_0(\bm{k}) \pm \Delta(\bm{k}),
\end{equation}
where $\Delta(\bm{k}) = |\bm{h}|=0$ over some 1D closed contour $\mathcal{N}$. 
This can then be gapped by introducing a term that breaks mirror symmetry, $f(\bm{k}) \sigma_y$, giving energy eigenstates:
\begin{equation}
    \ket{\pm} = \frac{1}{A_\pm} \left(
 \frac{\left(h_z\pm\sqrt{f^2+h_x^2+h_z^2}\right)}{h_x+i f} , 1 \right),
\end{equation}
where 
\begin{equation}
A_\pm = \sqrt{\frac{2 \sqrt{f^2 +h_x^2+h_z^2}}{\sqrt{f^2+h_x^2+h_z^2}\mp h_z}},
\end{equation}
is a normalization.
The nodal surface $\mathcal{N}$ naturally induces a set of 2D coordinates, $\left\{\bm{\kappa}(\bm{k}_0)\right\}$ at each point $\bm{k}_0 \in \mathcal{N}$, that is perpendicular to $\mathcal{N}$ at $\bm{k}_0$.
The band-resolved BC contributions along the direction of the nodal line are given by:
\begin{equation}
    \omega_\pm (\bm{k}_0) = -\mathfrak{Im}\left\{\bra{\nabla_{\bm{\kappa}}\pm}\times\ket{\nabla_{\bm{\kappa}}\pm}\right\}.
\end{equation}
Noting that over $\mathcal{N}$, $h_x =0, h_z =0$, such that $\ket{\partial_f \pm} = 0$, 
\begin{widetext}
    
\begin{equation}
\begin{split}
    \omega_\pm (\bm{k}_0) &= - \left[\mathfrak{Im}\left\{\varepsilon_{ijk} \left(    \frac{\partial h_x}{\partial \kappa_i}\frac{\partial\bra{\pm}}{\partial h_x} +\frac{\partial h_z}{\partial \kappa_i}\frac{\partial\bra{\pm}}{\partial h_z}\right)\left(    \frac{\partial h_x}{\partial \kappa_j}\frac{\partial\ket{\pm}}{\partial h_x} +\frac{\partial h_z}{\partial \kappa_j}\frac{\partial\ket{\pm}}{\partial h_z}\right)\right\}\right]_{\bm{k}=\bm{k}_0}\\
    &= -\left[2\mathfrak{Im}\left\{ \braket{\frac{\partial \pm}{\partial h_x}}{\frac{\partial \pm}{\partial h_z}}\right\}\left(\frac{\partial h_x}{\partial \kappa_x}\frac{\partial h_z}{\partial \kappa_z}-\frac{\partial h_x}{\partial \kappa_z}\frac{\partial h_z}{\partial \kappa_x}\right)\right]_{\bm{k}=\bm{k}_0}\\
    &= -\left[2\mathfrak{Im}\left\{ \braket{\frac{\partial \pm}{\partial h_x}}{\frac{\partial \pm}{\partial h_z}}\right\} \det\left[ J_{\bm{\kappa}}(\bm{h})\right]\right]_{\bm{k}=\bm{k}_0}\\
    &= \pm\frac{1}{2}\frac{1}{f(\bm{k}_0)^2}\det\left[ J_{\bm{\kappa}}(\bm{h(\bm{k}_0}))\right],\\
\end{split}
\end{equation}
\end{widetext}
where $J_{\bm{\kappa}}$ is the Jacobian.
Hence, unless $f(\bm{k})=0$ for some discrete $\bm{k}_0$, {\it i.e.} the gap is not open across the entirety of $\mathcal{N}$ (splitting $\mathcal{N}$ into pairs of Weyl nodes~\cite{Fang_2016,Huang,Weng_2015}), the sign of $\omega_\pm$ is determined by the sign of $\det[J_{\bm{\kappa}}]$. 
Crucially, given that $\mathcal{N}$ is, by definition, a level curve of $\bm{h}$, the derivative of $\bm{h}$ along $\bm{k}_0$ is zero, hence $\partial_{\bm{k}_0} J_{\bm{\kappa}} =0$, i.e the Jacobian is constant on $\mathcal{N}$. 
Hence, for a fully gapped nodal line with a simple 2D Hilbert space, $\omega_\pm$ always circulates around the nodal line with equal magnitude and opposite chiralities (or are zero).

For a nodal line with a richer 3D Hilbert space, we can always make a change of basis, $\ket{n} \rightarrow \ket{n'} = \mathcal{U}^\dagger(\bm{k}) \ket{n}$, to cast the Hamiltonian in the form
\begin{equation}
    \begin{pmatrix}
        \varepsilon_0(\bm{k}) + h_z(\bm{k}) & h_x(\bm{k}) & 0 \\
        h_x(\bm{k}) & \varepsilon_0(\bm{k}) + h_z(\bm{k}) & 0 \\
        0 & 0 & \zeta (\bm{k})
    \end{pmatrix}
\end{equation}
where the upper-left $2\times2$ block has the structure of a two-band nodal line and $\zeta(\bm{k})$ describes the dispersion of the `disconnected' band that does not directly take part in the nodal line crossings. 
In a trivial case where the nodal line and disconnected band do not couple over $k$-space, this sub-space is clearly separable, and we recover the result for the simple two-state case.
However, in the case of a nodal line with changing orbital/spin textures due to the presence of the third band, the $\bm{k}$ dependence of $\mathcal{U}(\bm{k})$ is not trivial.
This results in a band-resolved BC:
\begin{equation}
\begin{split}
\omega_n &= -\mathfrak{Im}\left\{\bra{\nabla_{\bm{\kappa}}n}\times\ket{\nabla_{\bm{\kappa}}n}\right\}\\
        &=  -\mathfrak{Im}\left\{\nabla_{\bm{\kappa}}\left(\bra{n'}\mathcal{U}^\dagger\right)\times\nabla_{\bm{\kappa}}\left( \mathcal{U} \ket{n'}\right)\right\}\\
&=-\mathfrak{Im}\left\{\bra{\nabla_{\bm{\kappa}}n'}\times\ket{\nabla_{\bm{\kappa}}n'}\right.\\
&\quad+\bra{\nabla_{\bm{\kappa}}n'}\mathcal{U}^\dagger\times\nabla_{\bm{\kappa}}\mathcal{U}\ket{n'}\\
&\quad+\bra{n'}\nabla_{\bm{\kappa}}\mathcal{U}^\dagger\times\mathcal{U}\ket{\nabla_{\bm{\kappa}}n'}\\
&\quad+\left.\bra{n'}\nabla_{\bm{\kappa}}\mathcal{U}^\dagger\times\nabla_{\bm{\kappa}}\mathcal{U}\ket{n'}\right\},
\end{split}
\end{equation}
where the first term captures the BC contributions of a two-state nodal line, and the $\mathcal{U}$-dependent terms capture the non-trivial evolution of the nodal line due to the effect of the third band. 
Due to the nature of solving generic $3\times3$ matrix eigenvalue problems, it is overly cumbersome to do further direct analysis to show explicitly the more complex BC textures that can arise.
However, the presence of the additional terms demonstrates that the circulation of BC is not as restricted as the two-state case. 

Therefore, we proceed numerically, introducing a mirror symmetry-breaking gap to the 3 Band toy model,
\begin{equation}
    \gamma \begin{pmatrix}
        0 & 1 & 0\\
        1 & 0 & 1\\
        0& 1 & 0
    \end{pmatrix},
\end{equation}
where $\gamma=0.001$~eV is the magnitude of the perturbation, the band resolved BC contributions are calculated in Fig.~\ref{fig:2}(d-f). The BC behaves in a more complex manner than what is possible in a two-state nodal line, with the critical points acting like sources and sinks that join opposite chirality $\omega_n$ flux lines. This supports the idea that the nodal lines are composite, formed from opposite chirality component nodal lines fused at nexus points.

\section{EuTiO$_3$}

%FIGURE3--------------
\begin{figure*}[ht]
    \includegraphics[width=\linewidth]{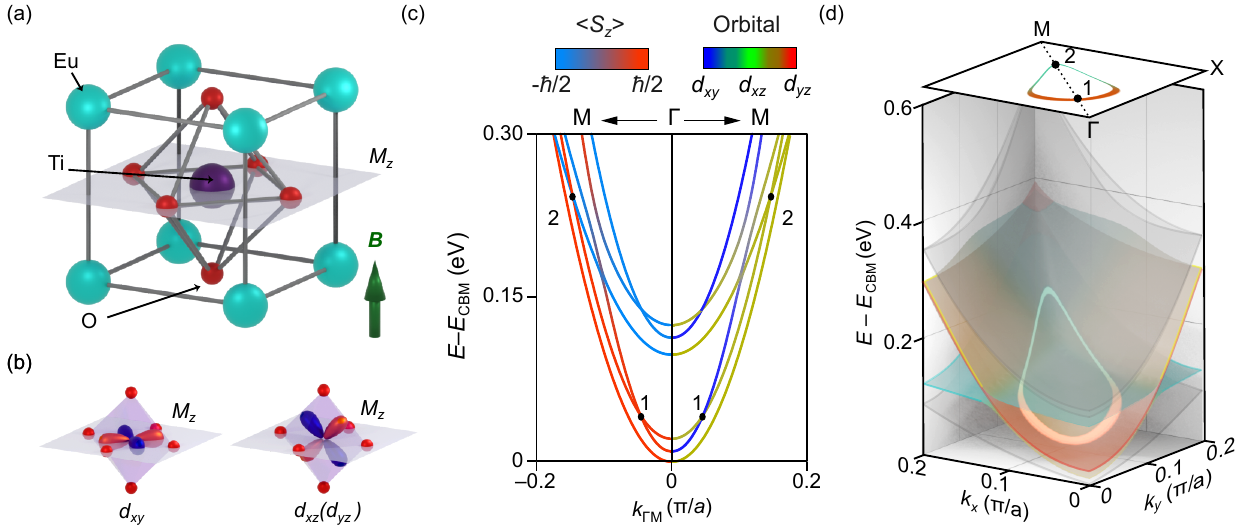}
    \caption{Overview of cubic EuTiO$_3$.
    (a) EuTiO$_3$ crystal structure, with a magnetic field, $\bm{B}$ along $z$ that leaves one mirror plane, $M_z$.
    (b) Schematic showing $d_{xy}(d_{xz},d_{yz})$ orbitals have a positive(negative) mirror eigenvalue.
    (c) EuTiO$_3$ conduction band structure along $\Gamma \rightarrow M$ high symmetry line, with $\langle \widehat{S}_z \rangle$ expectation value on the left side, and orbital projection on the right side. Protected crossings along this direction are marked as dots 1 and 2.
    (d) 3D view of $t_{2g}$ tight-binding band-structure in $k_z=0$ plane with the 3 bands involved in the composite nodal line color-coded red-yellow, and blue for the spin-up and spin-down respectively, and other bands are shown as grey.
    The composite nodal line is highlighted for clarity, with the inset showing its shape in $k$-space and dots corresponding to the intersection of the nodal line with $\Gamma \rightarrow M$ as shown in (c). The nodal line is color-coded for where the crossing occurs between bands of the same (red) and different (blue) spin.}
    \label{fig:3}
\end{figure*}
Moving beyond the simple toy model presented above, this section examines a realistic scenario for composite nodal line features in cubic, FFM EuTiO$_3$.
EuTiO$_3$ has a perovskite structure~\cite{Brous_1953} as depicted in Fig.~\ref{fig:3}(a), where the large magnetic moment of the Eu $4f$ electrons, $\sim 7 \mu_B$, contributes to an anti-ferromagnetic order below $5.5$~K~\cite{Lee_2009,Mcguire_1966,Chien1974,Gui2019}. Under a low magnetic field, $2.1$~T, EuTiO$_3$ transitions to an FFM phase by forcibly aligning the highly localised Eu $4f$ magnetic moments~\cite{Katsufuji_1999,Katsufuji_coupling_2001,Maruhashi_2020}. The RKKY interaction between the Eu $4f$ and Ti $3d$ states leads to a spin-polarization of the $3d$-$t_{2g}$ conduction bands along the magnetic axis, with the spin-up bands being brought down in energy by an effective Zeeman field.

In the case of an external magnetic field along (001), the original $O_h$ crystal symmetry is reduced to $C_{4h}$~\cite{Dowinton_2022,Maruhashi_2020}. This means all mirror symmetries are broken apart from reflection through the $xy$ plane, shown in Fig.~\ref{fig:3}(a), and it is left with a $4$-fold rotational symmetry about the $z$ axis.

Hence, it is useful to define the operator, $\widehat{R}$, that corresponds to the remaining mirror plane. The conduction bands are well described by a $6\times6$ Hamiltonian composed of $t_{2g}=\{d_{xz},d_{yz},d_{xy}\}$ orbitals with spin degree of freedom~\cite{Ranjan_2007,Maruhashi_2020,Dowinton_2022}.
A reflection through the $xy$ plane is equivalent to an inversion $\bm{r}\rightarrow\bm{-r}$ followed by a rotation $\pi$ about $z$, and it is naturally split into a spin and orbital component due to the highly spin-polarized conduction bands, {\it i.e.} $\widehat{R}=\widehat{R}_\sigma \otimes \widehat{R}_L$.

Inversion does not affect the spin degree of freedom hence $\widehat{R}_\sigma=e^{i \pi/2}\hat{\sigma}_z$ 
\begin{equation}
\begin{split}
\widehat{R}_\sigma\ket{\uparrow}&= i \ket{\uparrow}\\
\widehat{R}_\sigma \ket{\downarrow}&= -i \ket{\downarrow}.
\end{split}
\end{equation}
Likewise, for our $l=2$ spherical harmonics, inversion is irrelevant (as it contributes a phase $(-1)^l =1$), and we pick up a rotational phase $e^{im \pi}$. Such that,
\begin{equation}
\begin{split}
\widehat{R}_L\ket{d_{xz}}&=-\ket{d_{xz}}\\
\widehat{R}_L\ket{d_{yz}}&=-\ket{d_{yz}}\\
\widehat{R}_L\ket{d_{xy}}&=\ket{d_{xy}}.
\end{split}
\end{equation}
This is intuitive from the relationship between the real space $t_{2g}$ orbitals and the mirror plane, as shown in Fig.~\ref{fig:3}(b).

Topological Weyl semi-metal features have previously been connected to a non-monotonic anomalous Hall conductivity in EuTiO$_3$~\cite{Takahashi_2018,Ahadi2018,Dowinton_2022}.
To explore mirror protected Weyl nodal lines in FFM cubic EuTiO$_3$ a simple tight binding model is constructed. From symmetry considerations, the intra-orbital hoppings along $x$ for $d_{xy}, d_{xz}$ are equivalent and favoured over $d_{yz}$, and similarly for $y$ and $z$, hence there is a `perpendicular' and `parallel' mass, $m_{\perp}$ and $m_{\parallel}$.

For simplicity, it is assumed the magnetic field does not break this symmetry, only enhancing/suppressing hopping based on spin, $\sigma$, such that the kinetic energy for an orbital $d_{ij}$:
\begin{equation}
\begin{split}
    \varepsilon_{ij,\sigma}(\bm{k}) =& - \frac{1}{m_{\perp,\sigma}} \cos(k_l a)\\ 
        &- \frac{1}{m_{\parallel,\sigma}} (\cos(k_i a) + \cos(k_j a)),
\end{split}
\end{equation}
where $i,j,l \in \{x,y,z\}, i \neq j \neq l \neq i$.
Additionally, due to the exchange interaction of Ti $t_{2g}$ orbitals with Eu $4f$ there is an onsite Zeeman term: 
\begin{equation}
    \widehat{\mathcal{H}}_Z = \zeta \hat{\sigma}_z.
\end{equation}
While the crystal is inversion symmetric, and so there are no spin-orbit terms of the form $\bm{k}\times \bm{\nabla} V$, there are crucial `intrinsic' spin-orbit terms that mix our atomic orbital basis, including an onsite term:
\begin{equation}
    \widehat{\mathcal{H}}_\text{SO} = \alpha \hat{\bm{\sigma}}\cdot\widehat{\bm{L}}.
\end{equation}
Furthermore, there is a diagonal next nearest neighbour hopping term:
\begin{equation}
    \widehat{\mathcal{H}}_\text{SO,2} = \beta \left\{\widehat{L}_i,\widehat{L}_j\right\}\sin(k_i)\sin(k_j),
\end{equation}
where $\left\{\widehat{A},\widehat{B}\right\}= \widehat{A}\widehat{B} + \widehat{B}\widehat{A}$ is the anti-commutator.
This term couples different $t_{2g}$ orbitals without unquenching orbital angular momentum.
The parameters are chosen to reproduce \textit{ab-initio} DFT calculations and can be found in Appendix A.

Figure~\ref{fig:3}(c) shows the band-structure for the $6\times6$ tight binding Hamiltonian along the $\Gamma \rightarrow M$ high symmetry line, with orbital and spin expectation values on the right and left, respectively.
As discussed above, the spin-up $t_{2g}$ states are lower in energy due to the RKKY interaction with Eu $4f$ bands introducing an effective Zeeman field.
$\ket{d_{xz}}, \ket{d_{yz}}$ mix to form linear combinations with unquenched orbital angular momentum $\ket{d_{xz,yz}}$.
There is a crossing, 1, that takes place between $\ket{d_{xy},\uparrow}$ and $\ket{d_{xz,yz},\uparrow}$, which clearly must be protected due to mirror symmetry. 
For increasing $\bm{k}$ there is then an avoided crossing between the $\ket{d_{xy},\uparrow}$ derived band and $\ket{d_{xz,yz},\downarrow}$ derived band. Further along there is another crossing, $2$, between a $\ket{d_{xz,yz},\uparrow}$ derived band and $\ket{d_{xz,yz},\downarrow}$ derived which is again protected due to the differing mirror eigenvalues of the two bands. 

%FIGURE4--------------
\begin{figure*}[t]
    \includegraphics[width=\linewidth]{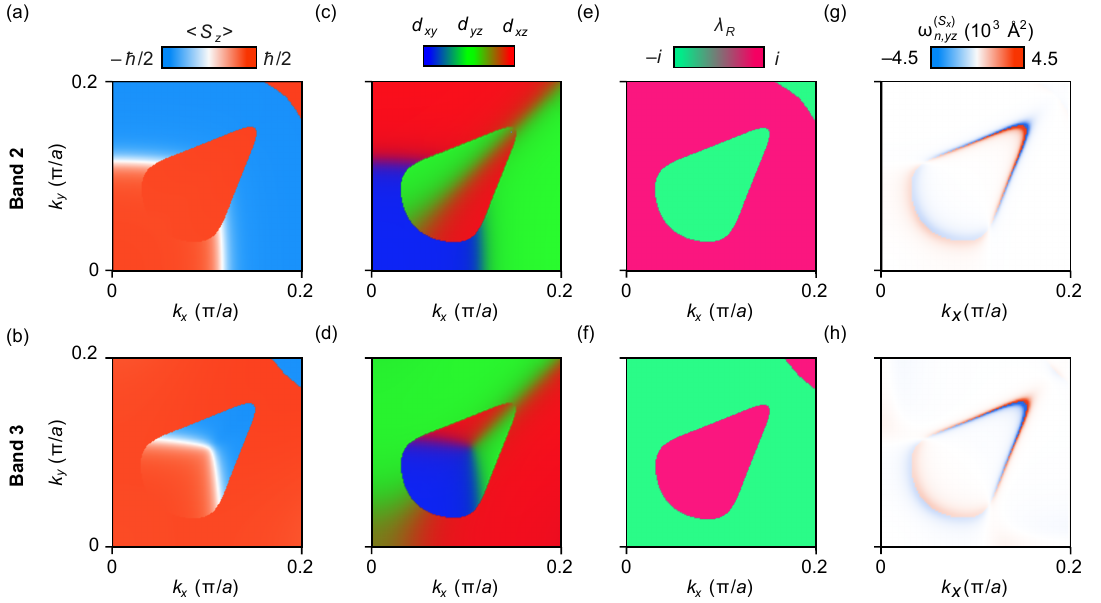}
    \caption{ (a-b) Spin $\langle \widehat{S}_z \rangle$, (c-d) orbital, (e-f) mirror textures, and (g-h) band resolved spin BC contribution, $\omega_{n,yz}^{(\widehat{S}_x)}$, in $k_z=0$ plane for $2^\text{nd}$ and $3^{\text{rd}}$ bands, ordered energetically, respectively.}
    \label{fig:4}
\end{figure*}
These mirror symmetry arguments apply to the whole $k_z =0$ plane, and hence, these crossings are part of nodal line features, not just 0D nodes. 
Figure~\ref{fig:3}(d) gives a 3D view of the band structure in the $k_z=0$ plane, with the three bands involved in the crossings shown in red and blue tones for spin up and down respectively, and the other three $t_{2g}$ bands greyed out. The nodal line is then highlighted, with red showing the region where the band crossing occurs between bands of the same spin (up) and cyan where the crossing is between bands of opposite spin. The inset shows the shape of this nodal line in $k$-space, and it is clear that the nodes 1 and 2 highlighted in Fig.~\ref{fig:3}(c), are the energy minima/maxima of two separate nodal lines that are then fused together due to the avoided crossing between the $\ket{d_{xy},\uparrow}$ derived band and $\ket{d_{xz,yz},\downarrow}$ band (seen intersecting $\Gamma \rightarrow M$ in Fig.~\ref{fig:3}(c).).

Figure~\ref{fig:4}(a-f) shows the spin, orbital, and mirror textures of the two bands, ordered energetically, involved in the nodal line crossing.
While the mirror eigenvalue remains quantized as required, the component spin and orbital degrees of freedom mix continuously due to the avoided crossing, giving a non-trivial evolution of spin and orbital textures over the region of the nodal line due to its composite nature. 

\section{Spin Hall Conductivity}
%FIGURE5--------------
\begin{figure}[t]
    \includegraphics[width=\linewidth]{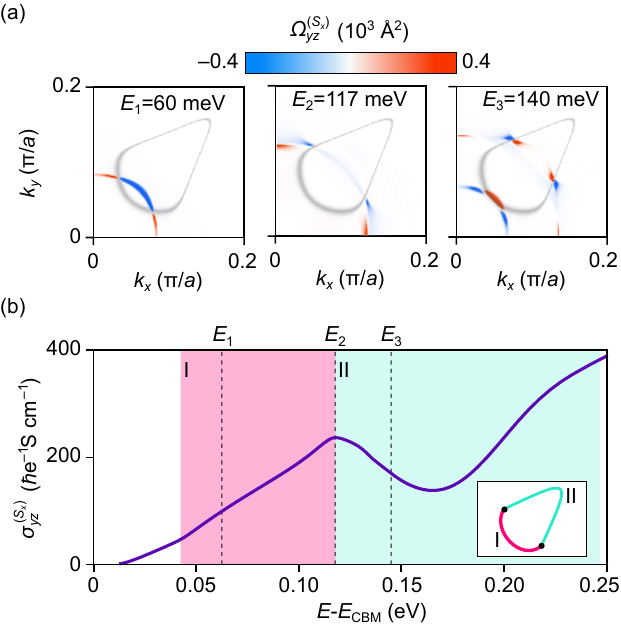}
    \caption{SHC in EuTiO$_3$. (a) Spin BC in $k_z =0$ plane at several energy cuts, $E_1 = 0.06$~eV $E_2 = 0.117$~eV $E_3 = 0.14$~eV. Grey shows the shape of a nodal line for clarity. (b) SHC calculated as a function of Fermi energy. Energies from (a) are marked for clarity. Regions I and II are color-coded to show the energy range of each half of the composite nodal line, with the locations in $k$-space shown schematically in the inset.}
    \label{fig:5}
\end{figure}
While the composite nodal line topology can be classified by the in-plane BC texture upon breaking mirror symmetry, this provides no experimental signature for un-gapped nodal lines when mirror symmetry is preserved. To develop an experimental signature of these nodal lines in EuTiO$_3$, and their non-trivial spin textures, it is worthwhile examining the SHC given by~\cite{Sinova2004,Guo2008,Gradhand_2012,Rauch_2017}
\begin{equation}
        \sigma_{ij}^{(\widehat{S}_l)} = -\frac{e^2}{\hbar}\int \frac{d^3 \bm{k}}{(2 \pi)^3} \Omega_{ij}^{(\widehat{S}_l)}(\bm{k}),
\end{equation}
where $\Omega_{ij}^{(\widehat{S}_l)}$ is a BC like term
\begin{equation}
    \begin{split}
        \Omega_{ij}^{(\widehat{S}_l)}(\bm{k})&=\sum_n f_n (T,\bm{k}) \omega_{n,ij}^{(\widehat{S}_l)}(\bm{k})\\
        \omega_{n,ij}^{(\widehat{S}_l)}&=-\hbar^2\mathfrak{Im}\sum_{m\neq n}\frac{\bra{n}\left\{\widehat{S}_l,\hat{v}_i\right\}\ket{m}\bra{m}\hat{v}_j\ket{n}}{|\varepsilon_n-\varepsilon_m-i\eta|^2},
    \end{split}
    \label{eq:SpinBC}
\end{equation}
with $\eta=0.0025$~eV being a smearing parameter. While Eq.~(\ref{eq:SpinBC}) is strictly speaking not a BC ~\cite{Gradhand_2012}, in our discussions, we will refer to it from here on as `spin BC' with the understanding that this nomenclature is for intuition and not precision.
With a $1/(\varepsilon_n-\varepsilon_m)^2$ dependence, Eq.~(\ref{eq:SpinBC}) is clearly sensitive to the nodal line crossing, and we can have a non-zero contribution in the $k_z=0$ mirror invariant plane by way of composing the mirror odd in-plane BC, $i=x,y$, $j=z$, with mirror odd in-plane spin, $l=x,y$.

Focusing on the $x (ij = yz)$ component, it is clear from the four-fold rotational symmetry of the system that the $l=y$ case has a vanishing contribution to linear SHC. Hence, our calculations focus on $\sigma_{yz}^{(\widehat{S}_x)}$.

Figure~\ref{fig:4}(g-h) shows that the nexuses of the composite nodal lines are critical points where the band-resolved contributions to the spin BC, $\omega_{yz}^{(\widehat{S}_x)}$ flip in character, analogous to the discussion for the regular BC contributions of gapped composite nodal lines above. Figure~\ref{fig:5}(a) shows the total spin BC for energies below ($E_1 =0.04$~eV), at ($E_2 = 0.117$~eV), and above ($E_3 = 0.14$~eV) the energy of the nexuses, at a temperature $\sim 5$K. Figure~\ref{fig:5}(b) shows the energy dependence of $\sigma_{yz}^{(\widehat{S}_x)}$, where the nexus of the composite nodal line appears as turning point, {\it i.e.} the SHC goes from increasing to decreasing with Fermi energy as the composite nodal lines shift in character at the nexus. Above a certain energy window, an additional nodal line between the spin-down bands (as can be seen at $E_3$) begins to dominate $\sigma_{yz}^{(\widehat{S}_x)}$. Hence,  $\sigma_{yz}^{(\widehat{S}_x)}$ is not a good experimental signature of the composite nodal line across its entire energy range, $0.04-0.245$~eV, but crucially is sensitive to the nexus points.
\section{Conclusion}
By intersecting three nodal lines in a singular mirror plane and then introducing a gap to one of them, composite nodal lines are formed from components of the original crossings fused at nexus points. While these nodal lines have a simple ring topology, it was shown they are capable of a more complex Berryology where the nexuses act like sources/sinks of BC. In addition, these composite nodal lines are expected to be realized in FFM cubic EuTiO$_3$, where this manifests as a nodal line with a simple ring topology but complex spin and orbital textures. A turning point in the linear SHC at the energy location of the nexuses provides a good experimental probe for this composite nodal line state. In the future, it could be interesting to explore non-linear SHC due to the nodal lines having multipole-like contributions to spin BC~\cite{Zhang_2023}.

\section{Methods}
Bulk electronic structure calculations for EuTiO$_3$ were performed within the density functional theory (DFT) using Perdew-Burke-Ernzerhof exchange-correlation functional~\cite{pbe} as implemented in the WIEN2K package~\cite{WIEN2K}. The relativistic effects, including spin-orbit coupling, were fully taken into account.  An effective Hubbard-like potential of $U_\text{eff} = 6$ eV for Eu was used to model the strong on-site Coulomb interaction of the Eu-$4f$ states~\cite{Maruhashi_2020}. 
A cubic crystal structure with a lattice parameter of 3.905 \AA~was used. 
Parameters for the tight binding model were then fit to DFT calculations to calculate electronic band structures and compute SHC as detailed in Guo et al. and Gradhand et al.~\cite{Guo2008,Gradhand_2012} A $k$-mesh of $576\times576\times576$, in $1/8$ of BZ from symmetry considerations, was used to compute the SHC.
\section{Acknowledgements}
The authors gratefully acknowledge the Research Infrastructures at the University of Manchester for providing the CSF3 high-performance computing facilities and the Center for Computational Materials Science at the Institute for Materials Research for allocations on the MASAMUNE-IMR supercomputer system (Project No. 202112-SCKXX-0510).  M.S.B. acknowledges support from Leverhulme Trust (Grant No. RPG-2023-253).
O.D. had support from EPSRC CDT Graphene NOWNANO, grant EP/L01548X/1. R.V.B. and M.S.B. are grateful to E-IMR center at the Institute for Materials Research, Tohoku University, for continuous support.
\appendix
\section{Tightbinding Model Parameters}
To reproduce DFT calculations, tight binding parameters found in \cref{tab:1} are used.
\begin{table}[h!]
\caption{Values of tight binding model parameters used to recreate DFT bands.}
\begin{ruledtabular}
 %   \centering
    \begin{tabular}{l|ll}
    Parameter & Value & Unit\\
    \hline  
        $m_{\perp,\uparrow}$ & 2.50 &eV$^{-1}$\\
        $m_{\parallel,\uparrow}$ & 0.667 &eV$^{-1}$\\
        $m_{\perp,\downarrow}$ & 14.9 &eV$^{-1}$\\
        $m_{\parallel,\downarrow}$ & 0.800 &eV$^{-1}$\\
        $\zeta$ & 0.049 &eV\\
        $\alpha$& 0.0122 &eV\\
        $\beta$&-0.05 &eV
    \end{tabular}
    
    \label{tab:1}
    \end{ruledtabular}
\end{table}
\section{Geometry of EuTiO$_3$ Nodal Lines}
This details a low $k$ model that is used to trace the geometry of band crossings relevant to the composite nodal lines in the $k_z=0$ plane.
\begin{equation}
\widehat{\mathcal{H}}=
\begin{pmatrix}
\widehat{\mathcal{H}}_\uparrow & 0 \\
0 &\widehat{\mathcal{H}}_\downarrow
\end{pmatrix}
\end{equation}
where each spin manifold has a Hamiltonian

\begin{widetext}
\begin{equation}
\widehat{\mathcal{H}}_\sigma =
\begin{pmatrix}
\zeta \lambda_\sigma+ m_\sigma(k^2 + \mu k_x^2) & -\beta k_x k_y - i \alpha \lambda_\sigma & 0 \\
-\beta k_x k_y +i \alpha \lambda_\sigma & \zeta \lambda_\sigma + m_\sigma(k^2 + \mu k_y^2) & 0 \\
0 & 0 &  \zeta \lambda_\sigma + m_\sigma(k^2 + \mu(k_x^2 +k_y^2))

\end{pmatrix}
\end{equation}
\end{widetext}
where $\zeta$ is a Zeeman term, $\alpha$ the onsite spin orbit term, $\beta$ the hopping dependent spin orbit term, and $\lambda_\uparrow = +1,\lambda_\downarrow=-1$ is the `sign' of spin.
$\mu,m_\sigma\geq0$ are terms that enhance/suppress the dispersion for specific orbitals or spin respectively, {\it i.e.} $d_{xy}$ is more dispersive along $k_x$ and $k_y$ than $k_z$ etc. 

For analytic simplicity, the $\alpha\hat{\sigma}_x \widehat{L}_x, \alpha\hat{\sigma}_y \widehat{L}_y$ spin-orbit terms are neglected, which is a reasonable approximation given that $|\zeta| >> |\alpha|$.
This model has energy eigenvalues
\begin{equation}
\begin{split}
    \varepsilon_{xy,\sigma} &= \zeta \lambda_\sigma + m_\sigma(k^2 + \mu(k_x^2 + k_y^2))\\
    \varepsilon_{\pm,\sigma} &=\zeta \lambda_\sigma + m_\sigma(k^2 + \frac{\mu}{2}(k^2 + k_z^2))\\
    &\pm \sqrt{k_x^2k_y^2\beta^2+\alpha^2+\frac{m_\sigma^2\mu^2}{4}(k_x^2-k_y^2)^2}.
    \label{eq:bands}
\end{split}
\end{equation}

%FIGURE6--------------
\begin{figure}[t]
    \centering
    \includegraphics[width=\linewidth]{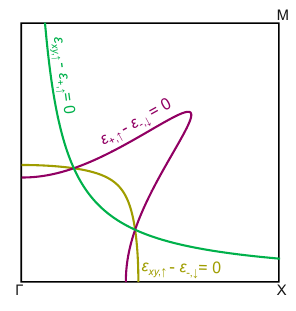}
    \caption{Geometry of $k_z=0$ crossings in EuTiO$_3$ taken from simple analytic model.}
    \label{fig:Appendix}
\end{figure}
From considering degeneracy in Eq.~(\ref{eq:bands}), the geometry of the nodal lines is constructed.
Firstly, examining the crossing between the $\ket{d_{xy},\uparrow}$ and $\frac{1}{\sqrt{2}}\left( i\ket{d_{xz},\uparrow}+\ket{d_{yz},\uparrow}\right)$ bands
\begin{equation}
\begin{split}
    \varepsilon_0 - \varepsilon_+ &= \frac{m_\sigma\mu}{2} (k_x^2 +k_y^2)\\
    &-\sqrt{k_x^2k_y^2 \beta^2 +\alpha^2 + \frac{m_\sigma^2\mu^2}{4} (k_x^2-k_y^2)^2}=0
\end{split}
\end{equation}
\begin{equation}
    \implies \frac{m_\sigma^2\mu^2}{4} (k_x^2 +k_y^2)^2 =k_x^2k_y^2 \beta^2+\alpha^2 + \frac{m_\sigma^2\mu^2}{4} (k_x^2-k_y^2)^2,
\end{equation}
which is satisfied by
\begin{equation}
    k_x^2 k_y^2 = \frac{\alpha^2}{m_\sigma^2\mu^2 -\beta^2},
\end{equation}
{\it i.e.} two hyperbolae.
We are also interested in the crossings between $\ket{d_{xy},\uparrow}$ and $\frac{1}{\sqrt{2}}(-i\ket{d_{xz},\downarrow}+\ket{d_{yz},\downarrow})$, and between $\frac{1}{\sqrt{2}}(-i\ket{d_{xz},\downarrow}+\ket{d_{yz},\downarrow})$ and $\frac{1}{\sqrt{2}}(i\ket{d_{xz},\uparrow}+\ket{d_{yz},\uparrow})$.
\begin{equation}
\begin{split}
\varepsilon_0^{(\uparrow)}-\varepsilon_-^{(\downarrow)}&=0\\
\varepsilon_+^{(\uparrow)}-\varepsilon_-^{(\downarrow)}&=0.
\label{eq:crossing23}
\end{split}
\end{equation}
These do not reduce down to a simple expression like the previous case, and hence, we solve them numerically; see Fig.~\ref{fig:Appendix}.
\bibliographystyle{apsrev4-1}

%\bibliography{refs.bib}

\end{document}